\documentstyle[12pt]{article}

\begin{document}

\begin{center}
{\large {\bf A G\"{o}del-Friedman cosmology?}}
\end{center}

\begin{center}
{\large Saulo Carneiro}\footnote{E-mail: saulo@fis.ufba.br}
\end{center}

\begin{center}
{\it Instituto de F\'{\i}sica, Universidade Federal da Bahia\\40210-340,
Salvador, BA, Brasil}
\end{center}

\begin{abstract}
Based on the mathematical similarity between the Friedman open metric and
G\"{o}del's one in the case of nearby distances, we investigate a new scenario
for universe's evolution, where the present Friedman universe is originated
from a primordial G\"{o}del's one by a phase transition during which the
cosmological constant vanishes. Using Hubble's constant and the present
matter density as input, we show that the radius and density of the
primordial G\"{o}del's universe are close, in order of magnitude, to the
present values, and that the time of expansion coincides with the age of
universe in the standard Friedman model. Besides, the conservation of
angular momentum provides, in this context, a possible origin for the
rotation of galaxies, leading to a relation between masses and spins
corroborated by observational data.
\end{abstract}

\section{Introduction}

${ }$

In spite of the paradigmatic role that standard Friedman model
has played in cosmology, astronomical observation is not
sufficiently precise and rich to close definitely the question
of what metric describes universe evolution \cite{Burbidge}. In
fact, since the original Einstein static universe, many
cosmological models have been proposed along the years
\cite{Weinberg}, each one of them in accordance with some
particular aspect of observation.

An important aspect of universe evolution, from both theoretical
and observational points of view, is the global rotation,
present in the literature since the pioneer works of Gamow
\cite{Gamow} and G\"{o}del \cite {Godel}, the latter presenting a
stationary solution for Einstein equations of gravitation.
Evidences of cosmological rotation have been reported \cite
{Birch}-\cite{Kuhne}, but always being object of controversy.
Another difficulty is related to the fact ``that it is
impossible to combine pure rotation and expansion in a solution
of the general relativity field equations for a simple physical
matter source" \cite{KO}.

In this paper we discuss a new scenario for universe evolution,
in which the universe observed nowadays, described by a Friedman
open metric, is originated from a primordial G\"{o}del's universe,
through a phase transition during which the negative
cosmological constant, characteristic of G\"{o}del's solution,
vanishes. The possibility of such a scenario is suggested by a
mathematical similarity between the Friedman open metric and
G\"{o}del's one, when we consider nearby distances.

This scenario is in accordance with our paradigm of an expanding
homogeneous universe and, in addition, it solves some problems
of the standard model, as, for example, the presence of the
initial singularity. In fact, using as input the Hubble constant
and the observed matter density, we show that the radius and
density of the primordial G\"{o}del's universe are close to the
present values, while the time of expansion coincides with the
age of universe in the standard Friedman model.

As a consequence of the conservation of angular momentum during the
G\"{o}del-Friedman phase transition, we show that the primordial
global rotation can naturally explain the origin of rotation of galaxies and
clusters, a hypothesis corroborated by the observed dipole anisotropy in the
distribution of the rotation axes of galaxies and, even quantitatively, by
the empirical relation between masses and spins of multi-stellar objects.

Of course, this model should be seen only as a {\it possible}
scenario, among many others, for universe evolution. Within the
spirit referred in the first paragraph, our goal is only to
provide an additional theoretical alternative to understand the
observational data we have in hand.

\section{The G\"{o}del-Friedman universe}

${ }$

In spherical coordinates ($r=sh\chi$, $\theta$, $\phi$, $\eta$), the open metric of Friedman is given by \cite
{Landau}

\begin{equation}  \label{1}
ds_F^2=a^2(\eta)[d\eta^2-d\chi^2-sh^2\chi(d\theta^2+sin^2\theta\;d\phi^2)]
\end{equation}

\noindent while the G\"{o}del one, written in cylindrical coordinates ($\rho=sh\xi$, $y$, $\phi$, $\eta$), has the
form \cite{Godel}

\begin{equation}  \label{2}
ds_G^2=a^2[d\eta^2-d\xi^2-dy^2+(sh^4\xi-sh^2\xi)d\phi^2+2\sqrt{2}\;sh^2\xi
d\phi d\eta]
\end{equation}

For nearby distances, $\xi\ll1$, and we have

\begin{equation}  \label{3}
ds_G^2 \approx a^2[d\eta^2-d\xi^2-dy^2-sh^2\xi d\phi^2+2\sqrt{2}\;sh^2\xi
d\phi d\eta]
\end{equation}

From the transformations

\begin{equation}
\rho=r\;sin\theta
\end{equation}

\begin{equation}
y=r\;cos\theta
\end{equation}

\noindent relating spherical and cylindrical coordinates, it follows

\begin{equation}  \label{4}
sh\xi=sh\chi\;sin\theta
\end{equation}

\begin{equation}
y=sh\chi\;cos\theta
\end{equation}

By differentiation we have

\begin{equation}
d\xi = \frac{1}{ch\xi} (sh\chi\;cos\theta d\theta+ch\chi\;sin\theta d\chi)
\end{equation}

\begin{equation}
dy=-sh\chi\;sin\theta d\theta+ch\chi\;cos\theta d\chi
\end{equation}

\noindent So, by using

\begin{equation}
\frac{1}{ch^2\xi} = \frac{1}{1 + sh^2\xi} \approx 1 - sh^2\xi = 1 - sh^2\chi\;sin^2\theta
\end{equation}

\noindent we obtain

\begin{equation}  \label{5}
dy^2+d\xi^2\approx\;sh^2\chi d\theta^2+d\chi^2
\end{equation}

Substituting (\ref{4}) and (\ref{5}) into (\ref{3}), one obtains

\begin{equation}  \label{6}
ds_G^2\approx a^2[d\eta^2-d\chi^2-sh^2\chi(d\theta^2+sin^2\theta
d\phi^2)+2\sqrt2\;sh^2\chi\;sin^2\theta d\phi d\eta]
\end{equation}

Comparing (\ref{1}) and (\ref{6}), we see that, unless for the
non-diagonal term, they are formally equal. This identification
leads us to conjecture the following possible model for
universe's evolution: $ds^2=ds_G^2$ for $\eta
<\eta _0$; $ds^2=ds_F^2$ for $\eta \geq \eta _0$ and $\xi \ll 1$.

Initially, we have a G\"{o}del universe, described by metric
(\ref{2}) or, for nearby distances, by metric (\ref{6}). At
$\eta=\eta_0$ a phase transition occurs, and the cosmological
constant vanishes. Now, the metric is non-stationary and, for
nearby distances, it coincides with the open metric of Friedman.
In the next section, we shall see that this match can be
performed in such a way that the scale parameter $a(\eta)$
changes continuously.

Though in principle we could find different ways to explain the
phase transition from G\"{o}del's universe to the Friedman one, a
plausible mechanism is provided if the cosmological constant is
identified with the energy density of a scalar field $\phi$,
with a self-interaction potential given, for instance, by \cite
{FL}

\begin{equation}  \label{FL}
V(\phi) = a\phi^2 - b\phi^3 + c\phi^4
\end{equation}

\noindent with $a,b,c>0$.

It is easy to verify that, for $b>2\sqrt{ac}$, this potential presents a
local minimum at $\phi=0$, for which $V(\phi)=0$, and an absolute minimum at

\begin{equation}
\phi = \frac{3b}{8c}\left[1 + \left(1-\frac{32ac}{9b^2}\right)^{1/2}\right]
\end{equation}

\noindent for which $V(\phi)$ has a negative value.

Therefore, the desired phase transition can be achieved, for
instance, by a tunneling of
the scalar field from the absolute minimum, for which the potential (\ref{FL}%
) (and then the cosmological constant) has a negative value --
as should be in G\"{o}del's model -- to the local minimum, for which
the potential vanishes, leading to a null cosmological constant
-- as expected in the standard Friedman model. In section 6 we
will return to this point, particularly in what concerns the
probability of such a transition.

\section{Numerical predictions}

${ }$

For $\eta\geq\eta_0$, the open model predicts \cite{Landau}

\begin{equation}  \label{8}
a(\eta)=a_0(ch\eta-1)
\end{equation}

\begin{equation}  \label{9}
t(\eta)=\frac{a_0}{c}(sh\eta-\eta)
\end{equation}

\begin{equation}  \label{10}
\rho a^3(\eta)=\frac{3c^2}{4\pi k}a_0
\end{equation}

\noindent where $a_0$ is a constant, $\rho$ is the matter density,
$k$ is the gravitational constant and we have introduced the
light velocity $c$.

From (\ref{8}) and (\ref{10}), it follows

\begin{equation}  \label{11}
a=a_0(ch\eta_0-1)=\frac{4\pi k\rho_G}{3c^2}a^3(ch\eta_0-1)
\end{equation}

\noindent which, by using the G\"{o}del radius \cite{Godel} $a^2=c^2(2\pi k\rho_G)^{-1}$, leads to $%
ch\eta_0=5/2$, or $\eta_0=1.6$, whatever the values of $a$ and
$\rho_G$, the matter density in the G\"{o}del phase.

Equations (\ref{8})-(\ref{10}) give

\begin{equation}  \label{12}
ch\left(\frac{\eta}{2}\right)=H_0\left(\frac{3}{8\pi k\rho}\right)^{\frac{1}{%
2}}
\end{equation}

\begin{equation}  \label{14}
a(\eta)=\frac{c}{H_0} coth\left(\frac{\eta}{2}\right)
\end{equation}

\begin{equation}  \label{15}
t(\eta)=\frac{1}{%
H_0}\left[\frac{coth(\eta/2)(sh\eta-\eta)}{ch\eta-1}\right]
\end{equation}

\noindent where $H_0 \equiv \dot{a}(\eta)/a^2(\eta) = h \times 100$ km/(sMpc)
is the Hubble constant. Using $h = 0.75$, $k=6.7 \times
10^{-11}$ m$^3$/kg.s$^2$ and $\rho=1.0\times10^{-27}$ kg/m$^3$
($\Omega = 0.1$), we have, for the present values of $\eta$,
$a(\eta)$ and $t(\eta)$, the values $\eta=3.7$, $%
a(\eta)=1.3 \times 10^{26}$m and $t(\eta)=3.8
\times 10^{17}$s.

Now, from (\ref{8}) we obtain

\begin{equation}  \label{13}
a(\eta)=a\left(\frac{ch\eta-1}{ch\eta_0-1}\right)
\end{equation}

\noindent which gives $a= 1.1
\times 10^{25}$m.

Moreover, (\ref{9}) leads to

\begin{equation}  \label{16}
t(\eta_0)=t(\eta)\left(\frac{sh\eta_0-\eta_0}{sh\eta-\eta}\right)
\end{equation}

\noindent i.e., to a time of expansion given by $\Delta t \equiv t(\eta) - t(\eta_0) = 3.6 \times
10^{17}$s.

For the open model, (\ref{10}) gives $\rho a^3(\eta)=\mbox{constant}$.
Therefore, $\rho a^3(\eta)=\rho_G a^3$, which leads to $\rho_G= 1.7 \times
10^{-24}$kg/m$^3$.

Finally, this last result leads us to an angular velocity for G\"{o}del's
primordial universe given by \cite{Godel} $\omega_G = 2(\pi k \rho_G)^{1/2}
= 3.8 \times 10^{-17}$s$^{-1}$.

Therefore, by using only two observed parameters as input, namely the
values for the Hubble constant and for the present matter density, we
conclude that the values for G\"{o}del's radius and density are close to the
present values, and that the time of expansion coincides with the age of
universe in the standard model.

It is important to verify how sensitive these results are respect
to the used value for the matter density of the present universe. If we use,
instead of $\Omega = 0.1$, the value $\rho=1.0%
\times10^{-28}$kg/m$^3$ ($\Omega = 0.01$), it is easy to obtain from the above formulae the
results $\eta=6.0$, $a(\eta)=1.3\times10^{26}$m, $a=1.0\times10^{24}$m, $%
t(\eta)=3.2\times10^{17}$s, $\Delta t=3.2\times10^{17}$s, $%
\rho_G=2.2\times10^{-22}$kg/m$^3$ and $\omega_G=4.3\times10^{-16}$s$^{-1}$.

\section{A possible origin for galaxies rotation}

${ }$

An important problem related to the G\"{o}del-Friedman phase transition is the
conservation of angular momentum: the Friedman model is isotropic, while in
G\"{o}del's one matter everywhere rotates relative to local gyroscopes. Now,
we will see that the conservation of angular momentum in the context of the
G\"{o}del-Friedman phase transition leads naturally to a possible explanation
for the origin of rotation of galaxies and clusters. For this purpose, we
shall adapt to our context a reasoning originally made by L. Li \cite{Li} to
describe galaxies formation in a rotating and expanding universe.

Let us consider, in the primordial G\"{o}del's universe, a proto-galaxy with
mass $M$, radius $r$ and density $\rho_G$, rotating with angular velocity $%
\omega_G$ with respect to a local inertial frame (a gyroscope frame).
Assuming for this proto-galaxy a spherically symmetric distribution of
matter, its angular momentum relative to the gyroscope frame will be given
by $J=2Mr^2\omega_G/5$, which can be rewritten, using $M=4\pi r^3\rho_G/3$,
as

\begin{equation}  \label{Li}
J=\frac{2}{5}\left(\frac{3}{4\pi\rho_G}\right)^{2/3}\omega_GM^{5/3}
\end{equation}

\noindent On the other hand, with respect to a galaxy frame whose origin is
fixed at the proto-galaxy center, the proto-galaxy angular momentum is zero
because, by definition, the galaxy frame co-rotates with the global rotation.

After the phase transition from G\"{o}del's universe to the Friedman one, the
galaxy frame does not rotate anymore with respect to gyroscope ones, and the
galaxy angular momentum relative to it will be given, due to the
conservation of angular momentum, by (\ref{Li}).

In this way, besides providing a possible origin for galaxies rotation, the
G\"{o}del-Friedman phase transition leads to a relation between angular
momentum and mass of multi-stellar objects corroborated by observational
data. In fact, the masses and spins of spiral galaxies, globular clusters
and clusters of galaxies (including the Local Supercluster)\footnote{%
See Table I of \cite{Muradian}.} can be fitted by the relation $J=\kappa
M^{5/3}$, with $\kappa\approx6.2\times10^{-2}$ (SI units)\footnote{%
If we use the relation $J=\kappa M^n$, with $n$ free, the fitting gives $n
\approx 1.69$ (while $5/3 \approx 1.67$) and $\kappa\approx0.48\times10^{-2}$
(SI units), in accordance with the value referred by Li \cite{Li}.}.

A theoretical value for $\kappa$ can be predicted from (\ref{Li}), but it
will depend on $\rho_G$ and $\omega_G$, i.e., on the value of the present
matter density used as input. For $\Omega = 0.1$ we have $%
\kappa\approx4.1\times10^{-2}$ and for $\Omega = 0.01$ we
obtain $\kappa\approx1.8\times10^{-2}$ (SI units).

If this picture for the origin of galaxies angular momentum is correct, an alignment of galaxies angular momenta along the
direction of global rotation of the primordial G\"{o}del's universe would be expected. This
prediction was already supported in the literature by the discovery of a
dipole anisotropy in the distribution of the rotation axes of galaxies \cite
{Li}, but, as pointed out by Li, an irregular shape of the proto-galaxies
can lead, after the phase transition, to an almost random distribution \cite
{Li}.

\section{The global metric}

${ }$

What happens for large distances after the phase transition? To
answer this question, let us write the G\"{o}del metric in the form

\begin{equation}
\label{global}
ds_G^2=a^2\{[d\eta + \Lambda sh^2(\xi/2)d\phi]^2
-(d\xi^2+dy^2+sh^2\xi d\phi^2)\}
\end{equation}

\noindent with $\Lambda = 2\sqrt{2}$. The $\Lambda$-term
originates the non-diagonal one in the nearby approximation
(\ref{6}), what suggests that the global metric arising from the
phase transition should be given by the non-stationary,
anisotropic metric

\begin{equation}
\label{anisotropic}
ds^2=a^2(\eta)[d\eta^2-(d\xi^2+dy^2+sh^2\xi d\phi^2)]
\end{equation}

This metric (which reduces to the Friedman open metric for
nearby distances) is solution of Einstein equations of
gravitation, with $\rho<\rho_c$, where $\rho_c = 3H_0^2/8\pi k$
is the critical density at the moment of observation. From it,
instead of relations (\ref{8})-(\ref{10}), we can obtain the
following ones

\begin{equation}  \label{8'}
a(\eta)= a_0 \left[ch\left(\frac{\eta}{\sqrt{3}}\right)-1\right]
\end{equation}

\begin{equation}  \label{9'}
t(\eta)=\frac{\sqrt{3}a_0}{c}\left[sh\left(\frac{\eta}{\sqrt{3}}\right)
-\frac{\eta}{\sqrt{3}}\right]
\end{equation}

\begin{equation}  \label{10'}
\rho a^3(\eta)=\frac{c^2}{4\pi k}a_0
\end{equation}

From them, it is easy to re-obtain the numerical predictions of section $3$.
For $\Omega=0.1$, we obtain $\eta_0=3.1$ (actually this value does not depends on $\rho$),
$\eta=6.6$, $a(\eta)=0.7\times10^{26}$m, $a=6.2\times10^{24}$m, $%
t(\eta)=3.5\times10^{17}$s, $\Delta t=3.3\times10^{17}$s, $%
\rho_G=1.4\times10^{-24}$kg/m$^3$ and $\omega_G=3.4\times10^{-17}$s$^{-1}$.
Besides, for the parameter $\kappa$ introduced in section $4$,
we obtain the value $\kappa=4.2\times10^{-2}$ (SI units). On the
other hand, for $\Omega=0.01$, we have $\eta=10.5$,
$a(\eta)=0.7\times10^{26}$m, $a=6.5\times10^{23}$m, $%
t(\eta)=4.0\times10^{17}$s, $\Delta t=4.0\times10^{17}$s, $%
\rho_G=1.3\times10^{-22}$kg/m$^3$, $\omega_G=3.3\times10^{-16}$s$^{-1}$
and $\kappa=2.0\times10^{-2}$ (SI units).

\section{Concluding remarks}

${ }$

We have seen that the mathematical similarity between the
Friedman open metric and G\"{o}del's one in the limit of nearby
distances suggests conjecturing about a new cosmological
scenario for universe's evolution, where the nowadays expanding
universe is originated from a primordial G\"{o}del's one. Besides
providing a possible origin for the rotation of galaxies, such a
cosmology solves some problems of the standard Friedman model,
as the presence of an initial singularity and problems related
to the age of universe: before the G\"{o}del-Friedman phase
transition, universe could have existed for a long time...

Of course, there remain several open questions. How to explain
hot universe phenomena, e.g. the cosmic microwave background
radiation or the nucleosynthesis, in the context of a
cosmological scenario where no dense phase exists? Despite the
possibility of explaining those phenomena without reference to a
hot phase \cite{Burbidge}, this question should be elucidated in
order to put this model in a more physical basis.

Another point which deserves better understanding is the
mechanism of phase transition. What is the probability of
tunneling for the scalar field? Is there inflation or the
production of bubbles and topological defects? Must the
cosmological constant be exactly zero after the phase
transition?

Actually, it is possible that all these problems are related,
and that a more detailed study of the phase transition mechanism
could shed light on problems like the origin of the cosmic
microwave background.

As discussed in section 2, a possible scenario for the phase
transition is provided by a tunneling of the scalar field from
the absolute minimum of potential (\ref{FL}) to the local one.
This kind of transition may be very improbable, but it is
precisely this fact that turns this scenario an interesting one.
Before the phase transition, we have a G\"{o}del, stationary
universe, where no cosmological time can be defined. During the
phase transition, we have a huge fluctuation with decrease of
entropy of the matter content, after which the universe starts
expanding and the cosmological time arrow coincides with the
thermodynamic one. Despite to be much improbable, this scenario
is not forbidden by any physical law, and the anthropic
principle could be used to justify it.

Another possibility is to consider (\ref{FL}) as an effective,
temperature dependent potential. In this case, the phase
transition can be caused by a change in the potential itself,
due to possible thermal processes in the primordial phase. Such
kind of transition of potential (\ref{FL}) was already
investigated in the scenario of QCD deconfinement \cite{QCD}.

In any case, a detailed study of the phase transition should
provide us a continuous match between the G\"{o}del metric and the
Friedman one, instead of the discontinuous match considered in
this paper, that could turn the model richer in several aspects.
A continuous transition from the absolute minimum of (\ref{FL})
to its local minimum necessarily crosses its local, positive
maximum, from which the potential rolls down to zero. This means
that in the latest phase of the transition a huge amount of
energy should be released, perhaps followed by an inflation
process, and that a reminiscent, positive cosmological constant
could exist nowadays, as suggested by recent observations
\cite{Supernova}. Besides, the necessity of a finite interval of
time for the transition to occur can lead to a primordial phase
more dense than the one obtained in this work.

Its also important to investigate the role of closed time-like
curves, characteristic of the G\"{o}del universe, in the process of
phase transition. Though we are considering, in the G\"{o}del phase,
only gravitational interaction between dust matter, the
appearance of CTC´s can occur during the phase transition
itself, where a self-interacting scalar field enters into play.
In any case, note that the eventual appearance of CTCs does not
affect the match considered in section 2, where only nearby
distances are considered. However, their presence can be
important when considering large ones, as done in section 5.

Finally, we would like to refer to the relation between this model and the ideas
concerning the similarity between the universe and elementary particles \cite
{SC}. If such similarity has a true physical significance, we could say that
the G\"{o}del-Friedman model describes a {\it decaying} universe. In this
context, the Dirac hypothesis \cite{Dirac} about the dependence of physical
quantities, as the gravitational constant, on the cosmological time is not
necessary anymore: the similarity should be found between particles (the
electron, in the Dirac original conjecture, or hadrons, in more recent ones)
and the primordial G\"{o}del's universe. In fact, using the values obtained
for $a$, $\rho_G$ and $\omega_G$ and typical values for the radius $r$, mass
$m$ and spin $\hbar$ of a particle, one can check the relations $a/r \sim
(M/m)^{1/2} \sim (J/\hbar)^{1/3} \sim 10^{40}$, where $M \sim \rho_G a^3$
and $J \sim M \omega_G a^2$ are, respectively, the mass and angular momentum
of the primordial universe.

\section*{Acknowledgements}

${ }$

I would like to thank P.F. Gonz\'{a}lez-D\'{\i}az and A. Saa for the
encouragement, R. Muradian, E. Radu and A.E. Santana for useful
discussions, and N. Andion for the reading of the manuscript.


\thebibliography{99}

\bibitem{Burbidge} G. Burbidge, IAU Symposium No. 183, Kyoto, Japan, 1997
(Kluwer Academic Publishers) [astro-ph/9711033].

\bibitem{Weinberg} S. Weinberg, 1972, {\it Gravitation and Cosmology} (John Wiley \& Sons),
chapter 16.

\bibitem{Gamow} G. Gamow, Nature 158(1946)549.

\bibitem{Godel} K. G\"{o}del, Rev.Mod.Phys. 21(1949)447.

\bibitem{Birch} P. Birch, Nature 298(1982)451; 301(1983)736.

\bibitem{NR} B. Nodland and J.P. Ralston, Phys.Rev.Lett. 78(1997)3043.

\bibitem{OKH} Y.N. Obukhov, V.A. Korotky and F.W. Hehl, astro-ph/9705243.

\bibitem{Kuhne} R.W. K\"uhne, Mod.Phys.Lett.A 12(1997)2473.

\bibitem{KO} V.A. Korotky and Y.N. Obukhov, {\it Gravity, Particles and Space-Time}, eds. P. Pronin and G. Sardanashvily (World Scientific: Singapore, 1996), pp. 421-439 [gr-qc/9604049].

\bibitem{Landau} L.D. Landau and E.M. Lifshitz, 1983, {\it The Classical Theory of Fields}
(Oxford: Pergamon), \S\S 111-114.

\bibitem{FL} R. Friedberg and T.D. Lee, Phys.Rev.D 15(1977)1694.

\bibitem{Li} L. Li, Gen.Rel.Grav. 30(1998)497.

\bibitem{Muradian} R. Muradian, Astrophys. Space Sci. 69(1980)339.

\bibitem{QCD} I. Bednarek, M. Biesiada and R. Manka,
astro-ph/9608053.

\bibitem{Supernova} P.M. Garnavich {\it et al.}, Astrophys.J. 509(1998)74.

\bibitem{SC} S. Carneiro, Found.Phys.Lett. 11(1998)95, and references therein.

\bibitem{Dirac} P.A.M. Dirac, Nature 139(1937)323.
\end{document}